\documentclass[12pt]{article}
\usepackage{array}
\usepackage{graphicx}
\usepackage{amssymb}
\usepackage{amsmath}
\usepackage{multirow}
\usepackage{cite}
\def\@fmsl@sh#1#2#3{\m@th\ooalign{$\hfil#1\mkern#2/\hfil$\crcr$#1#3$}}
 \def\eq#1\en{\begin{equation}#1\end{equation}}
\def\s[#1,#2]{[#1\stackrel{\star}{,}#2]}
\def\sx[#1,#2]{[#1\stackrel{\star_{x}}{,}#2]}

\newcommand{\nc}{\newcommand}
\nc{\beq}{\begin{equation}}
\nc{\eeq}{\end{equation}}
\nc{\beqa}{\begin{eqnarray}}
\nc{\eeqa}{\end{eqnarray}}

\def\bc{\begin{center}}
\def\ec{\end{center}}

\def\to{\rightarrow}

\def\gsim{\mathrel{\mathpalette\atversim>}}

\def\bc{\begin{center}}
\def\ec{\end{center}}

\def\gsim{\mathrel{\rlap{\lower4pt\hbox{\hskip1pt$\sim$}}

    \raise1pt\hbox{$>$}}}       

\def\gsim{\mathrel{\rlap{\lower4pt\hbox{\hskip1pt$\sim$}}
    \raise1pt\hbox{$>$}}}       

\begin{document}
\makeatletter
\def\fmslash{\@ifnextchar[{\fmsl@sh}{\fmsl@sh[0mu]}}
\def\fmsl@sh[#1]#2{%
  \mathchoice
    {\@fmsl@sh\displaystyle{#1}{#2}}%
    {\@fmsl@sh\textstyle{#1}{#2}}%
    {\@fmsl@sh\scriptstyle{#1}{#2}}%
    {\@fmsl@sh\scriptscriptstyle{#1}{#2}}}
\def\@fmsl@sh#1#2#3{\m@th\ooalign{$\hfil#1\mkern#2/\hfil$\crcr$#1#3$}}
\makeatother

\thispagestyle{empty}
\begin{titlepage}
\boldmath
\begin{center}
  \Large {\bf Gravitational Corrections to Fermion Masses in Grand Unified
Theories}
    \end{center}
\unboldmath
\vspace{0.2cm}
\begin{center}
{ {\large Xavier Calmet}\footnote{x.calmet@sussex.ac.uk} and {\large Ting-Cheng Yang}\footnote{ty33@sussex.ac.uk}
}
 \end{center}
\begin{center}
{\sl Physics and Astronomy, 
University of Sussex,   Falmer, Brighton, BN1 9QH, UK 
}
\end{center}
\vspace{\fill}
\begin{abstract}
\noindent
We reconsider quantum gravitational threshold effects to the unification of fermion masses in Grand Unified Theories.  We show that the running of the Planck mass can have a sizable effect on these thresholds which are thus much more important than naively expected. These corrections make any extrapolation from low energy measurements challenging. 
\end{abstract}  
\end{titlepage}



\newpage

There are several hints that the strong and electroweak forces unify at some very large energy typically assumed to be at around $10^{16}$ GeV. The quantum fields of the Standard Model fit nicely into simple representations of a Grand Unified Theory \cite{GeorgiGlashow} such as e.g. SU(5) or  SO(10). The idea of unification is extremely attractive for several reasons. For example, a grand unification drastically reduces the number of independent coupling constants. Furthermore, when extrapolated using renormalization group equations, the value of the strong and electroweak interactions measured at low energy seem to converge amazingly to some common value at around $10^{16}$ GeV \cite{Amaldi:1991cn} if the Standard Model is replaced by the Minimal Supersymmetric Standard Model at around a TeV. An important feature of Grand Unified Theories is that they predict the existence of many, potentially heavy, new particles. This is due to the very nature of Grand Unified Theories which need to be based on groups large enough to incorporate the Standard Model SU(3)$\times$ SU(2) $\times$ U(1) groups. Besides having to be large as such, unified theories often incorporate multiplets with a large number of fields to obtain viable phenomenology. When the unified theory is supersymmetric, the number of fundamental fields is even larger. It has been argued that the LHC data could be used to reconstruct, using renormalization group techniques, the fundamental Grand Unified Theory, see e.g. \cite{Baer:2008xc}, or differentiate between different supersymmetry breaking patterns \cite{Baer:2008xc}. In \cite{Calmet:2008df,Calmet:2009hp} it has been shown that there are potentially sizable quantum gravitational corrections to the unification conditions for the gauge couplings of the Standard Model. The thresholds have been known for a while \cite{Hill:1983xh,Shafi,Hall:1992kq,Vayonakis:1993nn}, but it had not been realized that they could potentially be larger than the two-loop corrections \cite{Calmet:2008df}. The aim of this work is to show that this quantum gravity blur has a  similar effect on the unification conditions for the masses of the fermions in a grand unified framework.

An important consequence of the large number of fundamental fields mentioned above, which can easily reach 1000, is that the scale at which quantum gravitational  effects are expected to become large is not necessarily as expected some $10^{19}$ GeV but is given by the renormalized Planck mass:
\begin{eqnarray}
M(\mu)^{2}=M(0)^{2}-\frac{\mu^{2}}{12\pi}(N_{0}+N_{1/2}-4N_{1})
\end{eqnarray}
with $M(0)$ is the Planck mass at low energy, i.e. Newton's constant is given by $G = M (0)^{-2}$, and $N_{0}$, $N_{1/2}$
and $N_{1}$ are respectively the numbers of real scalar fields, Weyl spinors and spin one vector bosons.

If the strength of gravitational interactions is scale dependent, the scale $\mu_*$ at which quantum gravity effects are large is the one at which
\begin{equation}
\label{strong}
M (\mu_*) \sim \mu_*~.
\end{equation} 
It has been shown in \cite{Calmet:2008tn}  that the presence of a large number of fields can dramatically impact the value $\mu_*$.  In many Grand Unified models, the large number of fields can cause the true scale $\mu_*$ of quantum gravity to be significantly lower than the naive value $M_{{\rm Pl}} \sim 10^{19}\,{\rm~GeV}$. In fact, from the above equations, one finds
\begin{equation}
\mu_* = M_{{\rm Pl}}/\eta~,
\end{equation}
where, for a theory with $N \equiv N_0+N_{1/2}-4 N_1$,
\begin{equation}
\eta=\sqrt{1+\frac{N}{12\pi}}~.
\end{equation}

In \cite{Calmet:2008df}, quantum gravity effects have been shown to affect the unification of gauge couplings (see 
 \cite{Hill:1983xh,Shafi,Hall:1992kq,Vayonakis:1993nn,Rizzo:1984mk,Datta:1995as,Dasgupta:1995js,Huitu:1999eh,Tobe:2003yj,Chakrabortty:2008zk},for a non-exhaustive list of papers). The lowest order effective operators induced by a quantum theory of gravity are of dimension five, such as \cite{Hill:1983xh,Shafi}
\begin{eqnarray}
\label{dim5} 
\frac{c}{\hat{\mu}_*} {\rm Tr}\left(G_{\mu\nu} G^{\mu\nu} H\right)~,
\end{eqnarray}
where $G_{\mu\nu}$ is the Grand Unified Theory field strength and $H$ is a scalar multiplet. This operator is expected to be induced by strong non-perturbative effects at the scale of quantum gravity, so has coefficient $c \sim {\cal O}(1)$ and is suppressed by the reduced true Planck scale $\hat{\mu}_*=\mu_*/\sqrt{8\pi}=\hat{M}_{{\rm Pl}}/\eta$ with $\hat{M}_{{\rm Pl}}=2.43\times 10^{18}\,{\rm GeV}$. 

The importance of  gravitational effects were illustrated in  \cite{Calmet:2008df} using the  example of SUSY-SU(5). Operators similar to (\ref{dim5}) are present in all Grand Unified Theory models and an equivalent analysis applies. 

In SU(5) the multiplet $H$ in the adjoint representation acquires, upon symmetry breaking at the unification scale $M_X$, a vacuum expectation value $\left\langle H \right\rangle = M_X \left(2,2,2,-3,-3\right)/\sqrt{50\pi\alpha_G}$, where $\alpha_G$ is the value of the SU(5) gauge coupling at $M_X$. Inserted into the operator (\ref{dim5}), this modifies the gauge kinetic terms of SU(3)$\times$SU(2)$\times$U(1) below the scale $M_X$ to
\begin{equation}
\label{gaugekineticterm}
\begin{split}
-&\frac{1}{4} \left(1+\epsilon_1\right)F_{\mu\nu} F^{\mu\nu}_{{\rm U}(1)}
-\frac{1}{2}\left(1+\epsilon_2\right){\rm Tr}\left(F_{\mu\nu} F^{\mu\nu}_{{\rm SU}(2)}\right)\\
& -\frac{1}{2}\left(1+\epsilon_3\right){\rm Tr}\left(F_{\mu\nu} F^{\mu\nu}_{{\rm SU}(3)}\right)
\end{split}
\end{equation}
with
\begin{equation}
\label{epsilons}
\epsilon_1=\frac{\epsilon_2}{3}=-\frac{\epsilon_3}{2}=\frac{\sqrt{2}}{5\sqrt{\pi}}\frac{c\eta}{\sqrt{\alpha_G}}\frac{M_X}{\hat{M}_{{\rm Pl}}}~.
\end{equation}
After a finite field redefinition $A_{\mu}^{i} \to \left(1+\epsilon_i\right)^{1/2} A_{\mu}^{i}$ the kinetic terms have familiar form, and it is then the corresponding redefined coupling constants $g_i \to \left(1+\epsilon_i\right)^{-1/2} g_i$ that are observed at low energies and that obey the usual RG equations below $M_X$, whereas it is the \emph{original} coupling constants that need to meet at $M_X$ in order for unification to happen. In terms of the observable rescaled couplings, the unification condition therefore reads:
\begin{equation}
\label{boundarycondition}
\begin{split}
\alpha_G & = \left(1+\epsilon_1\right) \alpha_1(M_X)=\left(1+\epsilon_2\right) \alpha_2(M_X) \\
& = \left(1+\epsilon_3\right) \alpha_3(M_X)~.
\end{split}
\end{equation}

In was shown in \cite{Calmet:2008df} that the effects can be larger than the two loop effects considered in e.g. \cite{Amaldi:1991cn} and that it could either invalidate claims of a perfect unification SUSY-Standard Model or on the contrary help to unify models which apparently would not unify their gauge couplings.

In this work we point out that the same physical effect can have important implications for fermion masses. Again we will be using a simple SU(5) model to make our point more explicit, but our results can be trivially generalized to any Grand Unified Theory. One of the most interesting predictions of a Grand Unified Theory,  besides the unification of the gauge couplings at the unification scale, is the unification of some of the fermion masses at the unification scale. 
Fermion masses are generated by the Yukawa interactions. For example, in the simple SU(5) grand unification model with a Higgs in the  ${\bf 5}$ representation, one has 
\begin{eqnarray}
\mathcal{L} & = & \{G_{d}\bar{\Psi}^{c}_{jR}\Psi_{kL}^{j}H^{k}(5)+G_{u}\varepsilon_{jklmn}\bar{\Psi}^{c \,  jk}_{L}\Psi_{L}^{lm}H^{n}(5)\}+h.c.\\
 & = & -\frac{2M_{w}}{\sqrt{2}g_{2}}[G_{d}(\bar{d}d+\bar{e}e)+G_{u}8[\bar{u}u]]
 \end{eqnarray}
and one obtains
\begin{eqnarray} 
m_d(M_X)=m_e(M_X)= -\frac{2M_{w}}{\sqrt{2}g_{2}} G_{d}
 \end{eqnarray}
where $M_w$ is the $W$-boson mass, $g_2$ the SU(2) gauge coupling and $G_i$ are Yukawa couplings. This is one of the most exciting results of Grand Unified Theories, namely at the unification scale $M_X$ the masses of the down-type quarks are equal to the masses of the charged leptons, while the mass of the u-type quarks are not related to other parameters of the model. The up-type quark masses are given by $m_u(X)=  -\frac{16M_{w}}{\sqrt{2}g_{2}}G_{u}$ at the unification scale.

In analogy to (\ref{dim5}), there are also dimension five operators which can affect the fermions masses. They have been considered a while ago by Ellis and Gaillard \cite{Ellis:1979fg} (see also \cite{Panagiotakopoulos:1984wf})
\begin{eqnarray} \label{dim5fermionsgen}
\frac{c}{\hat \mu_\star} \bar{\Psi} \phi \Psi H +h.c.
 \end{eqnarray}
 where $\Psi$ are fermion fields, $\phi$ and $H$ some scalar bosons multiplets chosen in appropriate representations. In a simple SU(5) toy model with scalar fields in the  ${\bf 24}$ and  ${\bf 5}$ representations, one gets 
\begin{eqnarray} \label{dim5fermions}
{\cal O}_{5} & = & \frac{a_1}{\hat{\mu}_{\star}}\{\phi_{mn}\bar{f}^{mk}H_{k}^{l}\Psi_{l}^{n}\}\nonumber \\
 & + & \frac{a_2}{\hat{\mu}_{\star}}\{\phi_{mn}H^{mk}\bar{f^{l}}_{k}\Psi_{l}^{n}\}\nonumber \\
 & + & \frac{a_3}{\hat{\mu}_{\star}}\{\phi_{mn}\bar{f}^{mk}H_{k}^{l}\Psi_{l}^{n}\}\nonumber \\
 & + & \frac{a_4}{\hat{\mu}_{\star}}\{\phi_{mn}H^{mk}\bar{f^{l}}_{k}\Psi_{l}^{n}\}\nonumber \\
 & + & \frac{a_5}{\hat{\mu}_{\star}}\varepsilon^{mnpql}\{\Psi_{mn}\Psi_{pq}H_{k}\phi_{l}^{k}\}
 \label{eq:dim5 operator},
 \end{eqnarray}
  where $\Psi$ and $f$ are fermion fields in ${\bf 10}$ and ${\bf 5}$ respectively.
In SU(5), the value of the expectations value of $\phi(24)$ and $H(5)$ are fixed by the requirement that the Grand Unified Theory be broken at some $10^{16}$ GeV, i.e  $\langle \phi(24)\rangle \sim 10^{16}$ GeV and that the spontaneous symmetry breaking of the electroweak interactions takes place at the weak scale, i.e. $\langle H(5) \rangle = 246$ GeV.

These operators lead to a modification of the unification condition for the down-type quarks and their respective charged leptons. One finds
\begin{eqnarray}
m_{d}(M_X)[1+2(\zeta_{1}+\zeta_{2}+\zeta_{3}-\zeta_{4})] =
 m_{e}(M_X)[1+\frac{9}{2}(\zeta_{1}-\zeta_{2}-\zeta_{3}+\zeta_{4})]
  \end{eqnarray}
  with
  \begin{eqnarray}
\zeta_{i}=\frac{-2 \sqrt{2}}{5 G_{d} g_u} \frac{M_{X}}{\bar{M}_{Pl}} a_{i} \eta
 \end{eqnarray}
where $g_u$ is the unified coupling constant. We note that $u$-type quark masses do receive a correction due to one of these operators:
\begin{eqnarray}
 m_{u}(M_X)(1+\frac{3}{8}\zeta_{5}).
 \end{eqnarray}
Clearly since the scale $\hat \mu_\star$, i.e., the effective reduced Planck mass, is very poorly known and depends of the number of fields in the unified theory it is very difficult to argue that these quantum gravitational effects can be neglected. While in this simple SU(5) model $\eta$ is only equal to 0.74 as shown in \cite{Calmet:2009hp}  $\eta$ can easily be as large as 8 in SO(10) models.
 The running of the Planck mass has thus potentially a large impact on the splitting at the unification scale of the down type quarks and down type leptons. It is easy to evaluate the magnitude of the effect. One finds $\zeta_i \sim 10^{-2}  a_i/G_{e} \eta$, where we used $\alpha_u \sim 1/40$ and $M_X/\bar{M}_{Pl}\sim10^{-2}$. Even if the $a_i$ are as tiny as the corresponding Yukawa couplings, one can get a 10\% effect for Grand Unified Theories with a large matter content and thus large $\eta$. Once again we see that renormalization effects of the Planck mass can have sizable effects on the unification conditions of Grand Unified Theories.

There are several implications of these results. Without a precise knowledge of the quantum gravitational corrections, i.e. of the full theory of quantum gravity, it is very difficult to extrapolate from low energy measurements to check whether fermion masses unify or not. This casts some doubts concerning the feasibility of reconstructing the parameters of a Grand Unified Theory by using low energy measurements performed at the Large Hadron Collider.  On the other hand, these threshold effects can help to explain the low energy pattern of fermion masses and can revive models which naively would predict the incorrect pattern in the low energy regime.

As a summary, we have reconsidered quantum gravitational threshold effects studied a long time ago by Ellis and Gaillard. We have shown that the running of the Planck mass can have a sizable effect and that these threshold corrections are much more important than naively expected. This result is in line with our previous observations concerning the quantum gravitational threshold corrections to the unification of the coupling constants of the Standard Model.


\bigskip{}

\end{document}